# Magnetic Tower Outflows from a Radial Wire Array Z-pinch

S .V. Lebedev[1], A. Ciardi[1], D. J. Ampleford[1], S.N. Bland[1], S.C. Bott, J.P. Chittenden[1], G. N. Hall[1], J. Rapley[1], C.A. Jennings[1]
and
A. Frank[2,3], E. G. Blackman[2,3], T. Lery[4]

[1]The Blackett Laboratory, Imperial College, London, SW7 2BW, UK
[2]Department of Physics and Astronomy, University of Rochester, Rochester NY 14627-0171
[3]Laboratory for Laser Energetics, University of Rochester, Rochester NY 14627-0171
[4]Dublin Institute for Advanced Studies, Dublin, Ireland

**Abstract:** We present the first results of high energy density laboratory astrophysics experiments which explore the evolution of collimated outflows and jets driven by a toroidal magnetic field. The experiments are scalable to astrophysical flows in that critical dimensionless numbers such as the Mach number, the plasma beta and the magnetic Reynolds number are all in the astrophysically appropriate ranges. Our experiments use the MAGPIE pulsed power machine and allow us to explore the role of magnetic pressure in creating and collimating the outflow as well as showing the creation of a central jet within the broader outflow cavity. We show that currents flow along this jet and we observe its collimation to be enhanced by the additional hoop stresses associated with the generated toroidal field. Although at later times the jet column is observed to go unstable, the jet retains its collimation. We also present simulations of the magnetic jet evolution using our two-dimensional resistive magneto-hydrodynamic (MHD) laboratory code. We conclude with a discussion of the astrophysical relevance of the experiments and of the stability properties of the jet.

**1. Introduction**
Jets and collimated "bipolar" outflows are a ubiquitous phenomena in astrophysics occurring in environments as diverse as newly forming stars (Young Stellar Objects: YSOs, Reipurth & Bally 2001), dying solar type stars (Planetary Nebulae, PN Balick & Frank 2002) and supermassive black holes at the center of Active Galactic Nuclei (AGN, see Begelman, Blandford, Rees 1984). Jets also likely play a critical role in the formation of Gamma Ray Bursts (GRBs for a review see Piran 2005) and perhaps by association (e.g. Galama 1998; Stanek et al. 2003) supernova as well (e.g. Leblanc & Wilson 1970; Khoklov et al 1999; MacFadyen and Woosley 1999; Wheeler, Meier , Wilson 2002; Akiyama et al 2003; Blackman et al. 2004 ).
 The presence of common outflow structures in such a diverse set of phenomena suggests a common formation mechanism. In many environments which are known to host jets and collimated outflows observational studies have imposed strong constraints on the ability of radiation pressure to drive the outflows, much less collimate them. For example, the momentum injection rates of the jets/outflows in both YSOs (Lada 1985) and PNe (Burarrabal et al 2001) have been shown to be many of orders of magnitude larger than that which can be supplied by photons from the central source. Thus some process other than radiation pressure is required which can both launch and collimate many of the observed classes of jets and bipolar outflows.

Over the last two decades magnetic fields have been identified as the principle, universal agent for creating collimated outflows. When a magnetic field is anchored in a rapidly rotating object at the bottom of a gravitational potential, the field can act as a drive-belt, tapping rotational energy and launching plasma back up the potential well. This *magneto-rotational* scenario for jet launching has taken a variety of forms and has been explored by numerous authors over the years both analytically (Blandford & Payne 1982, Pudritz & Norman 1986, Pelitier & Pudritz 1992, Wardle & Konigl 1993 Shu et al 1994, Ustyugova 1999, 2000) and numerically (Uchida & Shibata 1985, Contopolus & Lovelace 1996, Oyed, Pudritz & Stone 1997, Goodson, Winglee & Bohm 1999, Kudoh, Matsumoto & Shibata 2002). While there are significant differences between many of these studies in terms of details of the MHD processes involved, they share the common feature that jets are assembled when rotational energy is tapped to create magnetic stresses which then drive plasma to escape velocities.

Many of the models also share the property that the resulting flow pattern is dominated by a strong toroidal magnetic field. Indeed in most cases outflow collimation occurs via hoop stresses from such a toroidal component of the field. Many models also assume that these toroidal fields are not initial conditions, but result from a winding of initially poloidal field lines. Such winding is expected to occur due to differential rotation (see Shibata 1999 for a review of this issue) which can be expected in accretion disks around non-magnetized objects or when stellar fields are tied to a surrounding disk. In Lynden-Bell 1996, and Lynden-Bell 2003 a series of heuristic magneto-static models were developed which explored the fundamental structure of such jets called *magnetic towers*. In these models the height of the tower was proportional to the number of turns of the differentially rotating disk in which the fields were anchored. This picture of jet formation in which toroidal fields generated by differential rotation drive an outflow has been confirmed in a number of simulations (Romanova et al. 1998; Ustyugova et al. 2000; Kato et al 2004a,b).

While great progress has been made in understanding the magneto-rotational models for jet formation including those cases which could be described as magnetic towers, many questions require further study. For example it remains unclear how the currents, which support the magnetic field, are distributed in the outflow structure. Whether the currents return along the bow shock structure swept-up by the MHD winds or they flow in the ambient medium to close at infinity. The stability properties of the outflows and jets also remain an open question. While analytic studies and 3-D MHD simulations have made great progress there remain concerns about the role of initial conditions and numerical resolution. The cross sections of density, temperature and other variables across the jet and/or outflow can produce significant differences in the stability properties. In addressing these and other issues relating to magnetically driven outflows the advent of laboratory astrophysics experiments may be of use.

Experiments and numerical modelling of astrophysically relevant supersonic jets, produced using high intensity lasers, have been performed by a number of authors (Farley et al 1999, Logory et al 2000, Shigemori et al 2000, Stone et al 2000, Foster et al 2002,

Mizuta et al 2002). Using conical wire arrays on pulsed power facilities (Z-pinch machine), laboratory jet experiments with dimensionless parameters similar to young stellar object jets were also produced (Lebedev et al 2002, Ciardi et al 2002). All of the laboratory jets were formed hydrodynamically, through converging conical flows, which were either shock or ablatively driven. These experiments have shown the importance of radiative cooling on the collimation of the jets and have also proven capable of addressing issues relevant to astrophysical outflows, such as the propagation of a jet through an ambient medium (Logory et al 2000, Stone et al 2000, Foster et al 2002, Ampleford 2005) and a cross wind in which wind ram pressure can drive shocks into the jet beam and deflect its path (Ampleford et al 2004, Lebedev et al. 2004, Ciardi et al 2004).

To obtain information relevant to the launching mechanisms of the jets the experimental set-up should incorporate a dynamically significant magnetic field, which is thought to operate near the launch region of astrophysical jets. A magnetically driven jet experiment with high magnetic Reynolds number was proposed by Ryutov et al 2002. Experiments with magnetically driven jets were recently performed by Hsu & Bellan 2002 and 2003. Their experiments include both toroidal and poloidal magnetic fields, and provide, among other things, useful insights into the stability of magnetically dominated jets. The experiments do represent an important step forward in demonstrating a proof of principle of magnetic jet formation from initial magnetic loops. However it is not clear whether the dimensionless parameters of interest were in the correct regime, in particular these jets have low (<<1) plasma $\beta$ ($\beta = P_p/P_B$ ratio of thermal pressure to magnetic pressure) and they appear to be adiabatic, which is not likely to be the case for at least for protostellar astrophysical jets.

In this paper we present new laboratory experiments on the formation of astrophysically relevant, radiatively cooled supersonic jets driven by toroidal magnetic field pressure. Our experiments show the creation of both wide angle outflows and narrow collimated jet structures in which the plasma $\beta$ is of the order of unity, but which is surrounded by a low beta cavity. The present paper is intended as the first of a series and its purpose is to introduce the experimental set-up (section II), present the first results (section III) and discuss their astrophysical significance (section IV). In future works we will present a more complete theoretical analysis of the experiments and further our exploration of the results across parameter space.

## 2. Initial Evolution and Experimental Set-up
Magnetically driven jets are produced in our experiments using radial wire arrays, which consist of a pair of concentric electrodes connected radially by thin metallic wires (Fig.1a). In the experiment a 1 MA current, rising over 240 ns, is applied to the array. Resistive heating of the wires rapidly (~ 20 ns) converts them into a two-component plasma structure which consists of a cold, resistive dense wire core surrounded by hot, highly conductive low density plasma. The first phase of the experiment, where background ambient plasma is formed by wire ablation, is shown schematically in Figure 1b. Currents flow preferentially in the ablated plasma (low resistivity) in the immediate vicinity of the wire cores (high resistivity). While the ablated plasma is accelerated axially by the Lorentz $J \times B$ force, the wire cores instead remain stationary and act as a plasma source, which is continuously supplied by ablation of the wires. This plasma fills the region above the radial array, forming an ambient medium into which the magnetic tower will eventually expand. Close to the wires resistive diffusion dominates over the hydrodynamic convection of the magnetic field (magnetic Reynolds number $Re_M$ <1). The magnetic field and thus the currents remain confined in the proximity of the wires, leading to a relatively high β in the background plasma. Injection of plasma into the upper regions will continue until the wires are fully ablated and stop acting as mass sources.

Previous experiments with different configurations of wire array z-pinches (Lebedev et al., 2001) have shown that the ablation rate of the wires increases with the magnitude of the global magnetic field, which in a radial wire array changes as 1/r. The ablation rate is highest close to the axis and at some moment in time a section of the wires near the central electrode will be fully ablated. The disappearance of parts of the wire cores means that the swept up plasma cannot be replenished, and the current path shown schematically in Fig 1b is no longer available. Wire breakage thus leads to development of a magnetic cavity in the background plasma, which is pushed by the rising toroidal field loops (Fig 1c). This is the beginning of the second phase of the experiment: the formation of a magnetically driven jet. The current is now forced to flow along the surface of the cavity and through the central region, where a dense jet-like plasma column develops (Fig.1c and 1d). The pressure of the toroidal magnetic field, associated with the current flowing in the plasma column, leads to radial and axial expansion of the magnetic tower and to the axial acceleration of the jet column. Furthermore, the confinement of the magnetic cavity is largely determined by the thermal pressure of the background ambient plasma.

In the present experiments the radial array was made with 16 tungsten wires (13μm diameter). The central electrode of the array had a diameter 4mm, and the outer electrode 70mm. The wires were positioned in the radial plane or inclined at an angle of ~$10^0$ to the axis to form an inverse cone. Dynamics of the system was investigated using a laser probing system (λ=532nm, pulse duration 0.4ns) with interferometer and several schlieren channels, time-resolved XUV and soft x-ray imaging. The interferometer provided two-dimensional data on the distribution of the electron density in the observed outflow system, while the schlieren diagnostic is sensitive to the gradients of the refractive index. Four-frame gated soft x-ray pinhole cameras (2ns gate with 9ns separation and 5ns gate with 30ns separation) were used to record two-dimensional time-resolved images of the outflow structure emission. Spatial resolution for the laser probing was ~0.1mm and for the soft x-ray imaging ~0.3mm. Differentially filtered time-integrated x-ray images were used to estimate characteristic temperature in the plasma of the central column of the observed jet.

**3. Results.**
**3.1. Parameters of the background plasma.**
As discussed in the previous section, the background ambient plasma is produced by wire ablation. The expected distribution of this plasma can be calculated using a rate of mass ablation given by the model described in Lebedev et al., 2001:

$$\frac{dm}{dt} = \frac{\mu_0 I^2}{4\pi V r} \qquad (1)$$

where $I$ is the total current through the wires and the parameter $V$ is the ablation velocity ($V$~100km/s), which is approximately equal to the characteristic flow velocity far from the wires. This rate of ablation was also used in the 2-D resistive MHD calculations (Ciardi et al., 2005) as the rate of material injection on the computational grid. The computed radial distribution of mass density at a time corresponding to the end of the first phase and at the start of the magnetic jet formation is shown in Fig. 2a and Fig. 2b. An analytical formula closely approximating such distribution of mass can be obtained using Equation 1 and assuming that plasma is moving in the axial direction with a velocity equal to the ablation velocity. Taking into account the time-of-flight in the z-direction required for the flow to reach position z, the density distribution is given by:

$$\rho(r,t) = \frac{\mu_0}{8\pi^2 r^2 V^2}[I(t-\frac{z}{V})]^2 \qquad (2)$$

Equation 2 is obtained in the limit of a very large number of wires and thus represents an azimuthally averaged density distribution, which is equivalent to the 2-D axis-symmetric geometry used in the MHD simulations. Both approaches give similar distribution of the plasma for radii larger than the radius of the inner electrode. In addition, the simulations predict that the region near the axis will be filled by denser plasma, as discussed later.

In the experiments, the spatial distribution of the background ambient plasma was measured using laser interferometry (Fig.3a). The presence of plasma along the path of the laser leads to a shift of the interference fringes with respect to their position measured prior to the experiment in the absence of any plasma. The fringe shift is proportional to the integral of the electron density along the path of the probing laser beam. Fig.3b shows measured contour corresponding to the fringe-shift of 1 fringe. Position of this contour agrees well with that calculated using Equation 2 and assuming that the average ion charge Z=10, which is needed to make the transition from mass density in to the electron density measured by the interferometer. The measured distribution of the electron density in the background plasma is also in agreement with the distribution obtained in the 2-D MHD simulations (Ciardi et al., 2005). The high density plasma that is seen on axis is formed by converging background plasma driven by radial gradients of the plasma thermal pressure and, to a smaller extent, by the pressure of the magnetic field advected with the plasma flow. The formation of this axial plasma column is in fact similar to the formation of standing conical shocks seen in jet experiments using conical wire arrays (Lebedev et al 2002). Radiative cooling plays an important role in determining the size and properties of this plasma column, since it removes the excess thermal energy and allows the plasma to collapse to higher density and smaller diameters. Since it is mostly the radial component of the converging flow that is thermalized as it collides on axis, we note that the axial velocity in the plasma column is comparable to that of the background medium. The size of the plasma column increases with decreasing rate of radiative cooling, which may be varied, for example, using different wire materials. The total mass in the plasma column does not depend on the rate of radiative cooling and in the simulations it is similar to that measured in the experiment. However, in simulations with high radiative losses, the size of the plasma column is ~ 0.3 mm, significantly smaller than the size measured in the experiments. Because of the imposed symmetry, 2D axis-symmetric simulations with converging flows are prone to overestimate the density on axis, in particular when the reduced rates of radiative cooling due to effects of radiation transport are not included. For simulations with reduced radiation losses it is possible to get agreement with the measured size of the plasma column, ~1.5 mm in diameter.

The computed electron density above the wires decreases with height. For example, at t = 190ns the characteristic electron density at radius 2mm (equal to the radius of the central electrode) changes from $n_e \sim 5\times10^{18}$ cm$^{-3}$ at z = 2 mm to $n_e \sim 2\times10^{17}$ cm$^{-3}$ at z = 10 mm. The calculated average ionisation level (Z ~ 10-15) and temperature (~ 20 eV) of the background plasma remain fairly constant in space and time, and the thermal pressure of the ambient plasma is largely determined by its density distribution. By changing the size of the inner electrode and/or the inclination angle of the wires, it is possible to partially alter the density distribution of the ambient plasma and therefore its effects on the collimation of the magnetic cavity and jet.

**3.2. Formation of magnetically driven jets.**
Full ablation of wires forms gaps 0.5 - 1 mm long in the wire cores near the central electrode and halts the injection there of newly ablated plasma. This triggers, at ~220 ns, the formation of a magnetic cavity and the launching of a jet driven by rising toroidal magnetic loops. The evolution of a laboratory magnetic tower jet is illustrated in Fig.4, which shows a time sequence of soft x-ray self-emission taken during the same experiment. Images in Fig.4a,b were obtained using two x-ray cameras, filtered to transmit emission in the 150-280 eV range, observing the same jet from two different azimuthal directions, separated by 135$^0$. The brightest emission originates from the plasma on axis which is pinched and accelerated by the magnetic field. Emission from the expanding walls of the magnetic cavity indicates the position of the return current path. The size of the magnetic cavity increases in time both radially and axially, as shown in Fig.5. The expansion in the axial direction occurs with a velocity ~200 km/s, while the radial expansion is a factor of ~ 4 slower (~50km/s). This differential expansion leads to the elongation of the cavity in the z-direction. Two factors contribute to the faster growth of the magnetic tower in the axial direction. First, the magnetic pressure due to the toroidal magnetic field present in the magnetic cavity is highest near the axis, decreasing as r$^{-2}$ with the distance r from the axis. This means that the force acting on the wall of the cavity in the radial direction decreases as the cavity expands. In contrast, the force acting in the axial direction near the tip of the magnetic tower remains the same providing that the current reaches the tip of the central column and no "leakage" of the current across the cavity occurs. Second, the axial growth of the magnetic tower jet proceeds into the ambient plasma which is moving in the same direction with velocity of ~100km/s (~ablation velocity). As a result, the decelerating force arising from accreting mass by the moving wall of the cavity is smaller for the expansion in the axial direction. In addition, the ambient density decreases faster in the axial direction and drops to zero at a distance z ~ tV$_{abl}$ from the wires. It is also important to notice that the diameter of the magnetic tower is essentially fixed at the base, where the outer wall of the cavity (return current path) connects to the wires. This is because the current remains anchored to the wires at large radii where the wire cores have not been fully ablated. On the time-scale of the growth of the magnetic cavity (~40ns) only a small additional length of the wire could be fully ablated and the gaps in the wires should not increase by more than ~1mm.

The evolution of the magnetic cavity is also seen on the laser shadow images (Fig.6). This diagnostic is primarily sensitive to the transverse gradients of the electron density. If the gradients are higher than some threshold, the probing laser beam is deflected out of the optical system and shadows will result in the corresponding parts of the image. For our diagnostic set-up the threshold corresponds to L dn$_e$/dx > 8$\times10^{19}$ cm$^{-3}$, where L is the distance along probing direction. Due to mass accretion, the walls of the cavity are expected to have high density gradients and the shadow images indeed show sharp boundaries with characteristic thickness of the wall of ~0.5mm. At early time (~230ns) it is evident that the wall of the magnetic cavity is formed by discrete number of current paths, corresponding to the number of wires in the radial array. Despite the relatively small number of the wires in this experiment (16), the overall symmetry of the magnetic tower is very high, indicating equal division of the return current between the poloidal current loops. With time the magnetic tower elongates in the axial direction and the discreteness of the current paths becomes less pronounced.

At late times, as seen for example in Fig.6c, the outer wall of the magnetic cavity opens up at the top. This could be related to the steep decrease of the ambient plasma density with axial position z. The corresponding decrease in the decelerating (snowplow) force produces a rapid acceleration of the tip of the magnetic cavity and stretching of the plasma in the wall could also lead to the disruption of the path for the return current. In addition the current path supporting the toroidal magnetic field in the cavity could also be disrupted by the development of instabilities in the central jet.

The development of these current driven (MHD) instabilities in the jet column is clearly seen in both laser probing and soft x-ray emission images. The characteristic diameter of the central plasma column seen in x-ray images (Fig.4) is ~ 1.5 mm, though immediately after formation at ~235 ns it is smaller, ~ 0.5 mm. The presence of typical m=0 and m=1 perturbations is seen from the very beginning, which is consistent with the presence of large currents in the plasma column. The dominant wavelength of the perturbations (~1mm) is comparable to the diameter of the column and thus the dominant modes have $kr_0$ ~5. The characteristic growth rate for these MHD instabilities can be written as

$$\gamma = \Gamma(kr_0) \cdot \frac{C_A}{r_0} \qquad (3)$$

where $C_A$ is the Alfven velocity, $r_0$ is the jet radius and $\Gamma(kr_0)$ is a numerical factor which depends on the $kr_0$ of the mode, distribution of the current with radius and it is slightly different for the m=0 and m=1 modes. For $kr_0$ ~5, $\Gamma$ ~ 2 (Pereira et al., 1984) for both modes if the current is concentrated on the surface of the central plasma jet. The corresponding growth time for a typical jet radius of ~ 0.75 mm is ~ 2.5 ns, and the instabilities indeed evolve on this time-scale. Importantly, the experiment shows that the instability does not destroy the jet on a time-scale ~ 20 times longer then the instability growth time, and the amplitude of the kink mode remains comparable with radius of the central jet column.

The instabilities in the central part of the jet are also observed in the laser probing images (Fig.6). At early times the shadow of the plasma column on the axis is seen along the whole length of the magnetic tower. At later time, from about 260ns, only parts of the central column are seen as shadows, while in other parts the probing laser beam propagates through the plasma without being deflected out of the optical system by the density gradients. This implies that the density gradients in those regions are below the threshold of $Ldn_e/dx = 8 \times 10^{19}$ cm$^{-3}$. This observation allows us to estimate the characteristic line density of electrons ($N_e = \pi R^2 n_{av}$) in the central part of the jet by assuming that $dn_e/dx \sim n_{av}/R$ and that L is equal to the radius of the jet, R ~ 0.75mm. This gives a threshold line density $N_e$ ~$1.5 \times 10^{18}$ cm$^{-1}$, meaning that in the regions non-transparent to the probing laser beam, the line density is actually larger than this estimate.

The soft x-ray images show that the jet-like column, despite the presence of MHD instabilities, does not change its radius considerably over ~40ns and the plasma column appears to be in quasi-equilibrium, suggesting that the plasma β is of the order of unity. The assumption that β ~ 1 allows us to estimate temperature of this plasma using the Bennett relation (Bennett 1934):

$$\mu_0 I^2 = 8\pi N_i (Z+1) kT \qquad (4)$$

where $N_i$ is the ion line density, $Z$ is the average ion charge, $T$ is the plasma temperature and $I$ is the current in jet-like plasma column.

The ion line density is calculated assuming that the jet is formed by the ablated mass coming from the gaps (~ 1 mm) that develop in the wires at the start of the magnetic tower inflation. The mass of 1mm length of 16x13μm diameter tungsten wires is $4 \times 10^{-5}$g, which corresponds to an ion line density of $1.6 \times 10^{17}$ cm$^{-1}$ for 0.8cm long central jet column. If all current (~ 1 MA at this time) is flowing through the central jet column, the temperature required for the Bennett equilibrium is equal T=$2 \times 10^4$/(1+Z) eV. Assuming equilibrium ionisation, this corresponds to the Bennett temperature of 480eV and a LTE Z ~ 45. It is known however, that in Z-pinches formed from high atomic number materials, the Bennett temperature calculated for the total pinch current typically exceeds the measured plasma temperature. For example, for tungsten wire array Z-pinches (Deeney et al. 1997 and Cuneo et al., 2001) with values of $I^2/N_i$ comparable to that in our experiment, the measured plasma temperature was ~300eV and 170eV, respectively. The Bennett temperatures calculated from formula (3) is a factor of ~ 4 higher for both experiments (Deeney et al. 1997, Cuneo et al., 2001), presumably indicating that some fraction of current remains at a large radius in the low density trailing plasma. Assumption that the same reduction could be applied to our experiments (implying the same reduction in the current confining the plasma) gives an estimate of the plasma temperature of ~ 120 eV and Z ~ 20 in the central jet column, which is in good agreement with the simulated values. We will use these and the computed values in the discussion of dimensionless parameters of the jets presented in Section 3.3. The time-integrated differentially filtered X-ray images of the central jet column currently available do not contradict the above estimate, and we are planning to conduct time-resolved measurements of the plasma temperature in future experiments.

The evolution of the magnetic tower jet requires the presence of current which supports the toroidal magnetic field inside the cavity. The increase of the magnetic flux due to expansion of the cavity or, at later time, disruption of the current path through the central plasma column (or the wall of the magnetic cavity) should lead to an increase of the voltage applied to the gaps in the wires through which the inflation of the magnetic tower started. As a result it becomes energetically favourable for the current at some stage to reconnect through these gaps and to flow radially from the central electrode to the remaining length of the wires. This process is facilitated in the experiment by the filling of the gaps by plasma expanding from the electrode and from the wires. It is reasonable to assume that the characteristic expansion velocity of this plasma is similar to the typical velocity for the high current ion and electron beam diodes, ~ $2 \times 10^6$ cm/s. In this case the characteristic time for the filling of the gap is ~ 0.1cm / $2 \times 10^6$ cm/s ~ 50ns, which is comparable with the inflation time of the tower. After reconnection of the current the magnetic tower/plasma jet becomes detached from the source and will continue motion in the axial direction with velocity acquired during the stage of magnetic tower inflation. The detached jet could still have magnetic flux associated with the toroidal magnetic field and the decay of this magnetic flux will be determined by the value of the magnetic Reynolds number. For the plasma temperature in the central jet column of ~120 eV and Z~20, as estimated above, the magnetic Reynolds number $Re_M$ is ~10.

It is important to note that reconnection of current from the wires to the electrode restores the initial magnetic configuration, with the toroidal magnetic field concentrated in the region below the plane of the wires. The pressure of this toroidal magnetic field acting on the plasma filling the gap between the central electrode and the remaining length of the wires could drive the second episode of magnetic tower jet formation. A smaller amount of material is likely to participate in this process, suggesting a higher expansion velocity of the second magnetic cavity. The possibility of formation of the subsequent jet will be studied in the follow-up experiments.

**3.3 Dimensionless Parameters in the Experiment**
The relevance of the laboratory experiments to astrophysics rests on the capability of reproducing in laboratory both an adequate representation of the dynamics of the astrophysical system and, more importantly, a valid set of dimensionless parameters which are in the appropriate range of the astrophysical environment. The scaling conditions required in the hydrodynamic and magneto-hydrodynamic regimes can be found in Ryutov et al. 1999 and 200, where both the relevant dimensionless numbers and requirements to initial conditions are discussed.

In the present experiments, the evolution of a magnetic tower jet as it propagates through an ambient medium is studied. The effects of kinematic viscosity and thermal conduction are exemplified by the Reynolds (Re) and Peclet (Pe) numbers respectively. It is important to note that the dimensionless numbers associated with the ambient plasma and jet, are appropriate for astrophysics. The Reynolds number is large, $Re \sim 10^4$, and viscous effects can be neglected. The Peclet is not as large as in astrophysical systems but it is larger than 1 ($Pe \sim 5\text{-}20$). Thus we expect that heat conduction may play some role in smoothing out flow features. The plasma is also highly collisional, i.e. the ratio of the mean free path to the size of the system is small, m.f.p./jet radius ~ $10^{-5}$, and a fluid representation is therefore appropriate. Because magnetic fields are involved in the dynamics we also need to consider the plasma β and the magnetic Reynolds $R_M$ numbers. For the laboratory system the approximate calculated values are shown in Figure 7a which represents the snap-shot from simulations of the experiment corresponding to the time of the experimental image shown in Fig.7b. The current path is also schematically shown in Fig.7a. Near the wires, where the current follows a radial path, the magnetic Reynolds number is generally smaller than one ($Re_M \sim 0.1 - 0.5$). The effect of resistive diffusion is to confine the magnetic field mostly to the "foot" of the magnetic tower. Also in this region the plasma β is small (β ~ 0.1), and confinement of the magnetic cavity is partly due to the magnetic field pressure that is present in the ambient medium. The advected magnetic field in the background plasma, away form the foot of the magnetic cavity, is small and the plasma β increases with height until thermal pressure dominates over the magnetic pressure, β ~ 5. The magnetic tower is then confined by the thermal pressure of the ambient medium, and the collimation will depend on its density distribution (assuming that temperature and ionization are uniform). The magnetic

Reynolds number also increases with height. First, while the velocity of the plasma increases with z, the resistivity remains approximately constant. Second, the scale-length of the magnetic field gradients is larger for higher z, since there the magnetic field has a shallower profile. The resulting magnetic Reynolds number is $Re_M \sim 5$ and the characteristic diffusion time $\tau \sim 40\text{-}50$ ns is comparable to the magnetic tower evolution time. A $Re_M \sim 5\text{-}10$ is also calculated in the envelope of the magnetic cavity, where now the compression of the magnetic field leads to a plasma $\beta \sim 1$. Inside the envelope, in the magnetic cavity, the magnetic field dominates ($\beta \sim 0.01$) and the plasma is attached to the field lines ($Re_M \gg 10$). In the jet instead, there is balance between the thermal and magnetic pressures and the plasma $\beta \sim 1$, here again the advection of the field is dominant and $Re_M \sim 10$. As the magnetic cavity breaks up, the jet detaches and magnetic fields are entrained in the jet as it propagates upwards. The characteristic magnetic field diffusion time in the detached jet is $\tau > 100$ ns, which should allow the experimental detection of any currents present there.

Finally we give estimates of the characteristic Mach numbers (*M)* and cooling parameter ( $\chi$ ). The latter is defined as:

$$\chi = \frac{\varepsilon_T}{P_R \tau_H} \qquad (5)$$

where $P_R$ is the power radiated per unit volume, $\varepsilon_T$ is the thermal energy density and $\tau_H$ is a characteristic hydrodynamic time of jet evolution. In radiatively efficient regimes $\chi < 1$ and cooling through radiation is important in the dynamics and energy balance of the system. Assuming a characteristic time $\tau_H \sim 30-40$ ns and a radiated power calculated using the cooling tables given in Post et .al 1977, we estimate a cooling parameter in the range $\chi \sim 10^{-4} - 10^{-3}$, indicating that the jet is radiatively cooled. For the characteristic plasma parameters discussed at the end of Section 3.2, the internal Mach number of the jet on axis is estimated to be $M \sim 3 - 5$. Furthermore, the magnetic cavity is moving supersonically with respect to the ambient medium, which is also moving axially with a characteristic velocity of $\sim 100$ km s$^{-1}$. The jet Mach number for the motion of the magnetic cavity in the axial direction is $M \sim 10$, while in the radial direction $M \sim 3$.

A summary of the main physical and dimensionless parameters for the jet and ambient medium are shown in Table 1. Because of the somewhat broad variation of the parameters with position in the ambient medium some of those are not summarized in the table and the reader should instead refer to the text and to Fig. 7a.

**4. Discussion and Astrophysical Implications**
**4.1 Summary of the experimental jet formation and propagation**
The evolution of a magnetic jet produced in radial wire arrays proceeds in roughly three stages. During the first stage, background plasma is produced by wire ablation which continues until the wires are fully ablated. Due to the dependence of the ablation rate on the magnetic field, a gap in the wire cores initially develops near the central electrode and plasma ceases to be produced there. The start of the second stage occurs when the JxB force finally accelerates all of the plasma left along the gap in the wires, pushing the plasma upwards and sideways and forming a magnetic cavity. The subsequent evolution of the magnetic tower is dominated by the toroidal magnetic field present inside the cavity. The jet that develops on axis is part of a current path comprising the remaining wires and surrounding plasma, and the boundary of the magnetic cavity. The jet is in quasi-equilibrium (β ~ 1) and it is collimated by the magnetic hoop stress. As the magnetic tower elongates and inflates the current which it supports is roughly constant. The magnetic flux in the cavity increases and the magnetic tower inflates further. Collimation is provided by the thermal pressure of the background plasma (β > 1) which decreases both radially and axially. Furthermore, a high rate of radiative cooling aids the collimation the plasma.

The third stage of the experiment starts as the magnetic cavity breaks up, and the initial magnetic field configuration is, at least partially, restored. As the magnetic tower elongates, it becomes energetically favourable for the current to reconnect from the central electrode to the remaining wire sections. Three main factors can concurrently produce the detachment of the jet. Instabilities in the jet disrupt the current path along the jet column, the break up of the magnetic cavity envelope and the increase in the magnetic flux inside the cavity. As seen in Figure 6c and schematically in Figure 6d, a "knotty" jet is seen on axis after the break up of the magnetic cavity. If the detached jet maintains a high Reynolds number ($Re_M \gg 1$), so that the toroidal magnetic flux is conserved, currents will continue to circulate along poloidal loops. In the future experiments we are planning to test if the magnetic Reynolds number in the detached jet remains high enough to allow detection of entrained magnetic fields.

**4.2 Magnetically mediated astrophysical jet models**
Before summarizing the principles exemplified by the experiment, it is instructive to distinguish three related types of magnetically mediated astrophysical jet launch models and their relation to observed sources to which we can later refer:

*Steady magneto-centrifugal models (type A).* In these models, outflows are accelerated by a combination of centrifugal driving and magnetic pressure gradients (e.g. Ustyugova et al. 1998). Close to the rotator's (e.g. disc) surface, a strong poloidal field acts to enforce an approximate co-rotation. The centrifugal force then "flings" material along sufficiently inclined field lines (e.g. Blandford & Payne 1982; Pelletier & Pudritz 1992; Ouyed & Pudritz 2004). The total angular momentum per unit mass behaves as if the flow is rigidly rotating out to the Alfven radius, at which point the polodial outflow speed exceeds the Alfven speed associated with the poloidal field. Here the field incurs a significant lag behind the foot point motion and a toroidal field is generated. The toroidal field then finishes the job of accelerating the flow to its asymptotic speed, reached at scales less than ~100 times the initial foot point radius.

Such models comprise the leading paradigm for proto-stellar disk outflows. The actual launch region is not directly seen in YSOs observations as it is too small to be resolved. Recent observations of rotation in jets on ~ AU scales have provided new evidence for these models (Bacciotti et al. 2002; Coffey et al. 2004). On more easily observable scales, the collimated outflows appear to be super-Alfvenic (Reipurth and Bally 2001). We note that in the case where the initial poloidal field is weak, the winding of the field lines creates a strong toroidal field dominated jet even very close to the foot points (Kudoh & Shibata 1997, Shibata 1999).

In *transient magnetic explosion models (type B)*, an initial poloidal magnetic field is anchored in a spherical rotator, and the system is embedded in an overlying corona. The toroidal magnetic pressure gradient grows as the field is wound, due to the finite

Alfven propagation time in the presence of a significant density. When the toroidal field pressure gradient exceeds a critical value, the overlying material is blown off in a bipolar or quadrupolar outflow (Matt, Frank Blackman 2003). This is relevant for bipolar proto-planetary nebulae where an AGB star, with a rapidly rotating magnetized white dwarf core embedded in a diffuse envelope, may drive an asymmetric outflow. Analogous models have been studied with the goal of driving non-axisymmetric supernovae (e.g. Leblanc & Wilson 1970; Akiyama, et al 2003; Moiseenko et al. 2004; Blackman, Nordhaus, Thomas 2004). There the driving emanates ultimately from the field amplified from strong differential rotation near the neutron star core.

In both type A and type B jets, the asymptotic collimation can be a combination of integrated residual hoop stress over large propagation distances and the collimation of the initial launch direction in the sub-Alvenic regime. Both type A and B models are magnetically dominated only at the base, and become flow dominated outside the Alfven surface, typically a few engine radii from the launch point. This flow dominated regime occurs long before the jet comes to its end and thermalizes with the ambient medium.

By contrast, *Poynting flux dominated jets (type C)*, describes models in which the flow remains Poynting flux dominated over many decades in radius (e.g. Lovelace, Sulkanen, Wang 1987; Contopolous 1995; Lynden-Bell 1996; Romanova et al. 1998, Lyutikov & Blandford 2003), all the way to the dissipation region. Such models have been applied to relativistic jets of Active Galactic Nuclei and Gamma-ray bursts. Collimation of the Poynting flux "towers" requires an ambient plasma pressure (Lynden-Bell 1996).

Determining the scales out to which real jetted sources are magnetically or Poynting flux dominated is a topic of ongoing research, and is difficult without knowing the plasma composition. Predicting the distinguishing consequences of different instabilities in magnetically dominated vs. flow dominated jet plasma is a promising strategy (e.g Hardee 2004).

**4.3 Astrophysical jet principles exemplified by the experiment**
Our experiment, summarized in section 4.1, exhibits basic features that support the efficacy of magnetically mediated jet models of the type described in section 4.2. Roughly speaking, the structure we see at small heights from the launch surface resembles a broad magnetic tower reminiscent of Lynden-Bell (1996), but with magnetically pinched high beta plasma axial core. This geometry reveals the possibility that when the emission is proportional to positive power of the density, a high beta core appears as the observed jet even if it is surrounded by a low beta toroidal bubble. Note that at modest heights above the bubble, we see a ballistic jet resembling what is expected in the large distance regime of type A and type B models described above. However, the ballistic jet in the experiment is not driven by the bubble. Instead it was produced largely by the initial ablation of the wires and was thus present before the formation of the magnetic bubble

Presently, rather than trying to pinpoint the specific theoretical solutions corresponding to the experiment, we simply summarize the generic principles revealed:

(1) *Toroidal magnetic pressure gradients can drive outflows.* In astrophysical systems, the toroidal field is induced via radial differential rotation in the anchoring rotator and vertical shear due to a finite poloidal Alfven propagation time of the influence of the rotating base. In the experiment, the toroidal field is produced by the axial and radial currents and is initially confined below the wires until they are fully ablated near the electrode. Despite the difference in source of toroidal field, the experiment reveals that the toroidal field pressure gradient propels the plasma outward, just as in astrophysical jet models (e.g. Contopoulos 1995).

(2) *Hoop stress collimates flows.* The high density plasma on axis surrounded by the low beta, toroidal magnetic field dominated magnetic cavity reveals the importance of the magnetic hoop stress for collimation.

(3) *Two mechanisms of collimation: magnetic and non-magnetic.* Above the magnetically dominated cavity, there is a collimated ballistic flow. As mentioned above, this is a residual jet formed from the initial ablation of the wires and not the result of the magnetic cavity pressure. In a real jet of the type A and B described above however, it is possible that the magnetically dominated region would in fact directly transition to a collimated flow dominated regime not far from the source. The fact that we see a collimated jet at large height which is unrelated to the magnetic bubble exemplifies the principle that even when magnetic forces are not playing a strong role, a jet can be collimated if it is supersonic with an initially collimated launch direction and sufficient cooling [Lebedev et. al. 2002, Ciardi et. al. 2002]. That is, even if the jet is initially collimated by magnetic forces once it becomes ballistic and with high Mach number due to cooling, then the field no longer needs to exist to provide collimation.

(4) *Return current is carried by the magnetically dominated bubble.* A concern in MHD jet physics has been the location of the return current. That is, although the current flowing along the axis is understood to support the toroidal magnetic field there, where exactly does the return current flow? In the present experiment, the return current flows on the exterior of the magnetic bubble, far away from the high beta plasma jet core.

(5) *Time dependent dynamics are important.* By the very nature of the experimental setup, the system is transient, since the source of plasma is not steadily supplied once the wires ablate. There is no guarantee that a real astrophysical system should maintain a steady jet either. Pulsed jets have proposed to explain non-steady features in YSOs (Raga et al 2001, Gardiner & Frank 2000). As discussed above, given the finite supply of material from the ablated wires, it eventually becomes energetically favourable for the current to close inside the magnetic bubble. A series of such disconnected bubbles supplied by a non-steady outflow source in a real system could produce knots or could represent a single explosive outflow phase, analogous to the proto-planetary nebulae or supernovae driving concepts discussed above.

(6) *Instabilities need not destroy the overall jet flow.* Astrophysical outflows are expected to be liable to magnetohydrodynamic instabilities, which may affect their propagation, or even prevent them from maintaining their integrity over hundreds of kpc or several pc,. Various observed features such as wiggles, knots, and filamentary structures in extragalactic (e.g. Pearson 1996) and stellar (Reipurth & Heathcote, 1997) jets may on the other hand be manifestations of instabilities of the underlying flow. It has been a long standing question as to whether kink instabilities and/or Kelvin-Helmholtz instabilities occur in the different propagation regions of astrophysical jets and whether such instabilities destroy jets. This requires 3-D MHD studies (e.g. Ouyed & Pudritz 2003; Nakamura & Meier 2004). In our experiments, despite the presence of an instability, the basic structure of the jet remains intact (e.g. Fig. 5). The kink provides knots in the jet but not jet destruction, possibly suggesting the same for a real jet. Simulations of Ouyed & Pudritz 2003 and those of Nakamura et al. 2004 show that the Kelvin-Helmholtz and kink instabilities respectively, do not destroy the overall jet flow. The additional question of identifying which instability (KH vs. kink) operates in a real jet provides insight into whether the observed structure is magnetically confined or flow dominated in the vicinity of the observed jet. For example, from recent simulations (Hardee 2004) it appears that the kink instability propagates along the jet whereas the KH instability does not.

## 5. Conclusion

We have performed laboratory jet launching experiments that demonstrate how laboratory plasma experiments can provide insight into the physics of magnetically mediated astrophysical jet models. The experiments probe the launch region of the jet and reveal: a magnetically dominated cavity, produced by toroidal magnetic field pressure and a thermally dominated axial plasma core inside the magnetic cavity. This core is collimated by magnetic hoop stress and although later in the evolution, MHD instabilities disrupt the system, a clumpy radiatively cooled jet is still persistent. This indicates that instabilities need not be destructive. In the resulting clumpy jet there may be some entrained magnetic fields; however, given the high radiation losses and the resulting high Mach number of the flow, we expect its divergence to be small irrespective of whether magnetic confinement is still at play. Future experiments will probe the long term evolution of the clumpy jet and will try to determine the presence of any entrained magnetic field. Furthermore, in real accretion disks, the source of the toroidal magnetic field may result from the conversion of poloidal magnetic field due to differential rotation. In the present experiment no global poloidal fields are present and the toroidal field is the results of the current imposed on the radial wire array. The presence of a poloidal field could increase, for example, the stability of the jet. This possibility will be investigated in forthcoming experiments, where a poloidal field will be introduced in the system by rotating the wires around the central electrode, as in a solenoid.

**Acknowledgements**: This research was sponsored by the NNSA under DOE Cooperative Agreement DE-F03-02NA00057. Support to AF was also provided at the University of Rochester by NSF grant AST-0098442, and the Laboratory Astrophysics program at the Laboratory for Laser Energetics. The present work was supported in part by the European Community's Marie Curie Actions - Human resource and mobility within the JETSET network under contract MRTN-CT-2004 005592.

**Figures and Table captions**

**Figure 1. (a)** Schematic of a radial wire array experiment. Currents flow radially through fine metallic wires and along the central electrode, producing a toroidal magnetic field which lies below the wires. **(b)** The *J*x*B* force acting on the plasma ablated from the wires, produces a plasma background above the array, and because of resistive diffusion, the current path remains close to the wires. **(c)** Full ablation of the wires near the central electrode leads to formation of a magnetic cavity, which evolves **(d)** into a magnetic tower jet driven upwards by the pressure of the toroidal magnetic field.

**Figure 2.** Axis-symmetric, two-dimensional, resistive MHD simulations showing the spatial distribution of plasma mass density [kg m$^{-3}$] **(a)** before and **(b)** just after the beginning of magnetic cavity formation. The hatched areas indicate the position of the electrodes.

**Figure 3. (a)** Laser interferometry image of a region just above the wires. A higher density column is clearly seen on axis. **(b)** Measured contour corresponding to the fringe-shift of 1 fringe (dots and triangles are for the left and right-hand-sides of the image in (a)) and calculated using formula (2), as a function of axial and radial position.

**Figure 4.** Time sequence of soft X-ray images obtained during the same experiment showing expansion of the magnetic cavity and development of instabilities in the central jet column. The four images in part (b) were taken with smaller inter-frame time separations and from different viewing angle.

**Figure 5.** Axial and radial expansion of the magnetic cavity measured from images in Fig.4.

**Figure 6.** Laser shadowgraphs of the magnetic jet evolution. **(a)** At 233 ns the magnetic cavity is well developed. A collimated jet is clearly visible on axis inside the cavity. **(b)** The magnetic cavity elongates axially and expands radially. Because of instabilities sections of the jet on axis are no longer visible, and the jet assumes a clumpy structure. **(c)** The upper edge of the magnetic cavity breaks up and disappears. On axis is still visible a well collimated clumpy jet. **(d)** Schematic of the last stage of the magnetic jet evolution, showing how currents re-connect at the foot-point of the magnetic tower and a jet is ejected with entrained magnetic fields.

**Figure 7. (a)** Characteristic dimensionless parameters for a typical magnetic tower jet. **(b)** For comparison a soft-x-ray image of a magnetic tower jet is also shown.

**Table 1.** Summary of the physical and dimensionless parameters in the jet and in background plasma.